\documentclass[apjl]{ntemulateapj}     
\slugcomment{published in ApJL 2012 752 L13}
\usepackage{graphicx,amssymb,amsmath,times}
\usepackage{epsfig,epstopdf}
\usepackage[colorlinks=true,pdfstartview=FitV,linkcolor=blue,citecolor=blue,urlcolor=blue]{hyperref}  

\def\Journal#1#2#3#4{{#4}, {#1}, {#2}, #3} 

\newcommand{\etal}{\emph{et al.}}
\newcommand{\ie}{\emph{i.e.}}
\newcommand{\eg}{\emph{e.g.}}
\newcommand{\AMS}{\textsf{AMS}}
\newcommand{\Fermi}{\textit{Fermi}} 
\newcommand{\PAMELA}{\textit{PAMELA}} 
\newcommand{\ApJ}{ApJ}
\newcommand{\AeA}{A\&A}
\newcommand{\ApP}{APh}

\newcommand{\PRD}{PRD}
\newcommand{\PRL}{PRL}
\newcommand{\MNRAS}{MNRAS}

\newcommand{\JCAP}{JCAP}

\newcommand{\Hyd}{\textsf{H}}
\newcommand{\He}{\textsf{He}}
\newcommand{\Li}{\textsf{Li}}
\newcommand{\Be}{\textsf{Be}}
\newcommand{\B}{\textsf{B}}
\newcommand{\C}{\textsf{C}}
\newcommand{\N}{\textsf{N}}
\newcommand{\Oxy}{\textsf{O}}

\newcommand{\Fe}{\textsf{Fe}}
\newcommand{\BC}{\textsf{B/C}}
\newcommand{\pbar}{\ensuremath{\overline{\textsf{p}}}}
\newcommand{\pbarp}{\ensuremath{\overline{\textsf{p}}/\textsf{p}}}

\newcommand{\THM}{\textsf{THM}}
\newcommand{\THMa}{\textsf{THMa}}
\newcommand{\THMb}{\textsf{THMb}}
\newcommand{\HM}{\textsf{HM}}

\begin{document}

\title{Origin of the Cosmic Ray Spectral Hardening}
\shorttitle{Origin of the Cosmic Ray Spectral Hardening} 

\shortauthors{Nicola Tomassetti}
\author{Nicola Tomassetti}
\affil{\it INFN -- Sezione di Perugia, 06122 Perugia, Italy; \\e-mail: nicola.tomassetti@pg.infn.it}

\begin{abstract}  
Recent data from ATIC, CREAM and \PAMELA{} indicate that the cosmic ray
energy spectra of protons and nuclei exhibit a remarkable hardening at energies above 100\,GeV  per nucleon.
We propose that the hardening is an interstellar propagation effect that 
originates from a spatial change of the cosmic-ray transport properties in different regions of the Galaxy.
The key hypothesis is that the diffusion coefficient is not separable into energy and space
variables as usually assumed.
Under this scenario, we can reproduce the observational data well. Our model has several implications for the 
cosmic-ray acceleration/propagation physics and can be tested by
ongoing experiments such as the Alpha Magnetic Spectrometer or \Fermi/LAT. 
\end{abstract}

\section{Introduction}    
\label{Sec::Introduction} 
Understanding the spectral features in cosmic rays (CRs) is of fundamental importance for any theory of their origin and propagation.
The spectrum of CR nuclei above $\sim$\,10\,GeV per nucleon is thought to be the result of 
acceleration mechanisms in supernova remnants (SNR), which is steepened by propagation in the 
interstellar medium (ISM) due to leakage from the Galaxy \citep{Strong2007}.
The acceleration of primary CRs is described by the
diffusive shock acceleration (DSA) theory that predicts power-law
spectra $Q \propto E^{-\nu}$ with slope $\nu\approx$\,2.0--2.2.
The CR transport in the turbulent magnetic fields is modeled as a
diffusion process, with coefficient $K \propto E^{\delta}$, into a
magnetic \textit{halo} of typical height $2L\sim$\,10\,kpc.
Interactions of CRs with the ISM give rise to secondary nuclei such as
\Li-\Be-\B. 
This description predicts power-law spectra such as $\sim$$E^{-\nu -\delta}$ for primary nuclei (\eg, \Hyd, \He) 
and $\sim$$E^{-\delta}$ for secondary-to-primary ratios (\eg, \BC).
The parameters $\nu$ and $\delta$ are not firmly predicted \textit{a priori} and can be inferred from the data.
Observations constrain $\nu+\delta\approx$\,2.7 (depending on the element) and $\delta \sim$\,0.3--0.7,
whereby $\nu\sim$\,2.0--2.4. 
Yet, this picture dramatically over-predicts the CR anisotropy at $\gtrsim$\,TeV energies, which would suggest an almost
energy-independent diffusivity. 
Although the CR flux may also be affected by stochasticity effects on SNR events \citep{Blasi2012b},
the high degree of isotropy observed in CRs poses a
serious challenge to the conventional approach to the CR propagation.
Even more dramatically, the CR spectra at $E\gtrsim$\,100\,GeV/nucleon
exhibit a remarkable hardening with increasing energies. 
This feature has been established by recent experiments ATIC-2 \citep{Panov2009},
CREAM \citep{Ahn2010} and \PAMELA{} \citep{Adriani2011}, 
though the data are a little discrepant with each other. 
The proposed explanations 
interpret the hardening as a \textit{source effect} connected with 
acceleration mechanisms \citep{Biermann2010}, 
nearby SNRs \citep{Thoudam2012}
or arising by different populations of CR sources \citep{Zatsepin2006,Yuan2011}.
The subsequent impact of the hardening on the secondary CR production was estimated
by \citet{Lavalle2011} for $e^{+}$, \citet{Donato2011} for \pbar{} and $\gamma$--rays, 
and \citet{Vladimirov2011} for light nuclei. 
The latter also considered a possible diffusive origin of the effect, 
described by means of an effective break in the slope of $K(E)$,
around $\sim$\,200\,GeV/nucleon, from $\delta\cong$\,0.3 to $\delta\cong$\,0.15. 

Contrary to these works, we propose that the hardening originates from a 
spatial change of the CR diffusion properties in the different regions of the Galaxy.
In fact, propagation studies assume that CRs experience the same type of diffusion in the whole 
propagation region, with only few exceptions for particular environments (\eg, in SNR shells or 
inside the heliosphere) that are treated separately. 
A more realistic position-dependent diffusivity (correlated with the SNR density)
is considered in some works 
\citep[\eg,][]{Shibata2004,Gebauer2009,DiBernardo2010},
but they still adopt a unique energy dependence for $K$
in the whole halo, which leads to essentially unchanged results for the shapes of CR spectra at Earth.
Such descriptions may represent an over-simplification of the problem. 
In theoretical considerations, the diffusion is caused by the CR scattering on hydromagnetic waves which, 
in turn, depends on the nature and scale distribution of the magnetic-field irregularities. 
While SNR explosions may generate large irregularities in the region near the Galactic plane,
the situation in the outer halo is much different because there are no SNRs.
The main source of turbulent motion in the halo is presumably represented by CRs themselves.
From these considerations, \citet{Erlykin2002} found that the turbulence spectrum in the halo 
(in the far outer Galaxy) should be flatter than that in the plane (in the inner Galaxy). 
This implies strong latitudinal changes (gradual radial changes) for the
parameter $\delta$
and suggests spectral variations of CRs throughout the Galaxy. 
Noticeably, new data reported by \Fermi/LAT on the diffuse $\gamma$-ray
emission at $\sim$\,10--100\,GeV of energy seem to support these suggestions.
The $\gamma$-ray spectra observed near the Galactic plane (latitude
$|b|<$8$^{\circ}$) are found to be harder than those at higher latitudes, and
similar differences are also found between the inner Galaxy
(longitude $l<80^{\circ}$ or $l > 280^{\circ}$) and the outer Galaxy \citep{Ackermann2012}.

In this Letter, we focus our attention on the latitudinal changes of
the CR diffusion properties (which are expected to be more extreme) and
we show that they inevitably lead to a pronounced hardening for 
the energy spectra of CR nuclei at Earth.
Using analytical calculations, we examine the implications of this scenario for the main 
CR observables --primary spectra, secondary-to-primary ratios and latitudinal anisotropy--
and discuss its  connections with the open problems in the CR acceleration/propagation physics.

\section{Calculations}    
\label{Sec::Calculations} 

We use a simple model of CR diffusion and nuclear interactions. The effects of energy 
changes and convection are disregarded. The Galaxy is modeled to be a disk, with half-thickness $h$, 
containing the interstellar gas (number density $n$) and the CR sources. The disk is surrounded by a 
diffusive halo of half-thickness $L$ and zero matter density. 
For simplicity we give a one-dimensional description (infinite disk radius) in the
thin disk limit  ($h \ll L$) and we restrict to stable species. 
For each CR nucleus, the transport equation reads
\begin{equation}\label{Eq::Transport1D}
\frac{\partial N}{\partial t} = \frac{ \partial}{\partial z} \left( K(z) \frac{\partial N}{\partial z} \right) 
-2h \delta(z) \Gamma^{\rm inel} N + 2h\delta(z) Q \,,
\end{equation}
where $N(z)$ is its number density as function of the $z$-coordinate, $K(z)$ is the position-dependent diffusion 
coefficient, and $\Gamma^{\rm inel} = \beta c n \sigma^{\rm inel}$ is the destruction rate in the ISM 
at velocity $\beta c$ and cross section $\sigma^{\rm inel}$. 
The source term $Q$ can be split into a primary term $Q_{\rm pri}$, from SNRs, 
and a secondary production term $Q_{\rm sec}= \sum_{\rm j} \Gamma_{j}^{\rm spall} N_{\rm j}$,
from spallation of heavier ($j$) nuclei with rate $\Gamma_{j}^{\rm spall}$.
The quantities $N$, $K$, $Q$, and $\Gamma^{\rm inel}$ depend on energy
too. Since no energy changes are considered, such a dependence is only implicit and can be ignored for the moment.  
To solve Eq.\,\ref{Eq::Transport1D} we assume steady-state conditions ($\partial N/\partial t =0$). 
We define $\alpha_{1}(z) \equiv K^{\prime}/K$, $\alpha_{2}\equiv
-2h\Gamma^{\rm inel}/K_{0}$, and $\alpha_{3} \equiv 2hQ/K_{0}$, 
where we have denoted $K^{\prime}=\partial K / \partial z$ and $K_{0} = K(z$=$0)$. 
In the halo ($z \lessgtr 0$) the equation reads
$N^{\prime\prime}+\alpha_{1} N^{\prime}=0$, which is readily solved as $N_{\pm}(z) = p_{\pm} + u_{\pm} \hat{\lambda}(z)$,
where the subscripts $\mp$ indicate the solutions in the $z\lessgtr 0$ half-planes. The function $\hat{\lambda}(z)$ is defined as
\begin{equation}\label{Eq::LeffvsZ}
\hat{\lambda}(z) = \int_{0}^{z} e^{-\int_{0}^{z^{\prime}}\alpha_{1}(z^{\prime\prime})dz^{\prime\prime}}dz^{\prime} \,.
\end{equation}
The boundary conditions, $N(\pm L)=0$, provide the relation $u_{\pm}=-p/\Lambda_{\pm}$, 
where $\Lambda_{\pm} \equiv \hat{\lambda}(\pm L)$. 
From the continuity condition in the disk, $N_{-}(0) \equiv N_{+}(0)$, one obtains $p_{+} = p_{-}\equiv p$.
Assuming that $K(z)$ is an even function, one can see that $\hat{\lambda}(z)$ must be odd. 
Thus we define $\lambda(z)\equiv \hat{\lambda}(|z|)$ and $\Lambda\equiv \lambda(L)$, so that $u_{\pm}=\mp p/\Lambda$.
To determine $p$, we integrate the transport equation in a thin region across the disk
\begin{equation}\label{Eq::DiskIntegration}
  N^{\prime}_{+}(\epsilon) - N^{\prime}_{-}(-\epsilon) +
  \int_{-\epsilon}^{+\epsilon} \alpha_{1} N^{\prime} dz + \alpha_{2} N(0) + \alpha_{3} = 0 \,.
\end{equation}
The limit $\epsilon\rightarrow 0$ gives $p=\alpha_{3}/\left(2\Lambda^{-1} - \alpha_{2} \right)$. 
After replacing $\alpha_{1}$, $\alpha_{2}$, and $\alpha_{3}$ with the original quantities, 
the solution reads 
\begin{equation}\label{Eq::SolutionVSz}
  N(z)= \frac{Q}{\frac{K_{0}}{h\Lambda}+ \Gamma^{\rm inel}} \left[1 - \frac{\lambda(z)}{\Lambda} \right] \,.
\end{equation}
The function $\lambda(z)$ can be expressed as
\begin{equation}\label{Eq::Identity}
\lambda(z) = \int_{0}^{|z|} e^{-\int_{0}^{z^{\prime}} K^{\prime}/K dz} dz^{\prime} = K_{0} \int_{0}^{|z|} \frac{dz}{K(z)} \,.
\end{equation}
From this toy model, one can recover the homogeneous diffusion model 
(\HM) by setting $K^{\prime}\equiv0$, 
which gives $\lambda=|z|$ and $\Lambda=L$. 
Simple models of \textit{inhomogeneous diffusion} can be described by a diffusion coefficient of the type $K(z,E) \equiv f(z)K_{0}(E)$.
For example, \citet{DiBernardo2010} adopt 
$f(z)=e^{|z|/z_{t}}$.  
In this case one finds $\lambda= z_{t} \left( 1 - e^{-|z|/z_{t}} \right)$,
where the limit $z_{t} \gg L$ recovers the \HM, and the limit $z_{t}\ll L$ gives $\Lambda\approx z_{t}$. 
The latter case provides a more natural description 
of the latitudinal CR density profile, as it is insensitive to the halo boundaries $\pm L$.
However, the model predictions in terms of CR spectra at Earth remain equivalent to those 
of the \HM{} after a proper choice of $\Lambda$. 
This is a general property of Eq.\,\ref{Eq::Identity}: 
as long as $K(z,E)$ is separable in $z$ and $E$,
the function $\lambda$ is independent on energy and
the spectra at $z$=$0$ are equivalent to a mere rescaling of the model parameters. 
We remark that the energy--space variable separation is implicitly
assumed in all CR propagation models.
Physically, it describes a unique diffusion regime in the whole halo, given by $K_{0}(E)$,
while $f(z)$ allows for spatial variations in its normalization. 
The quantity $\Lambda$ can be regarded as an \textit{effective halo height} experienced by CRs at equilibrium.
\\

\begin{figure*}
\begin{center}
\epsscale{1.15}
\plotone{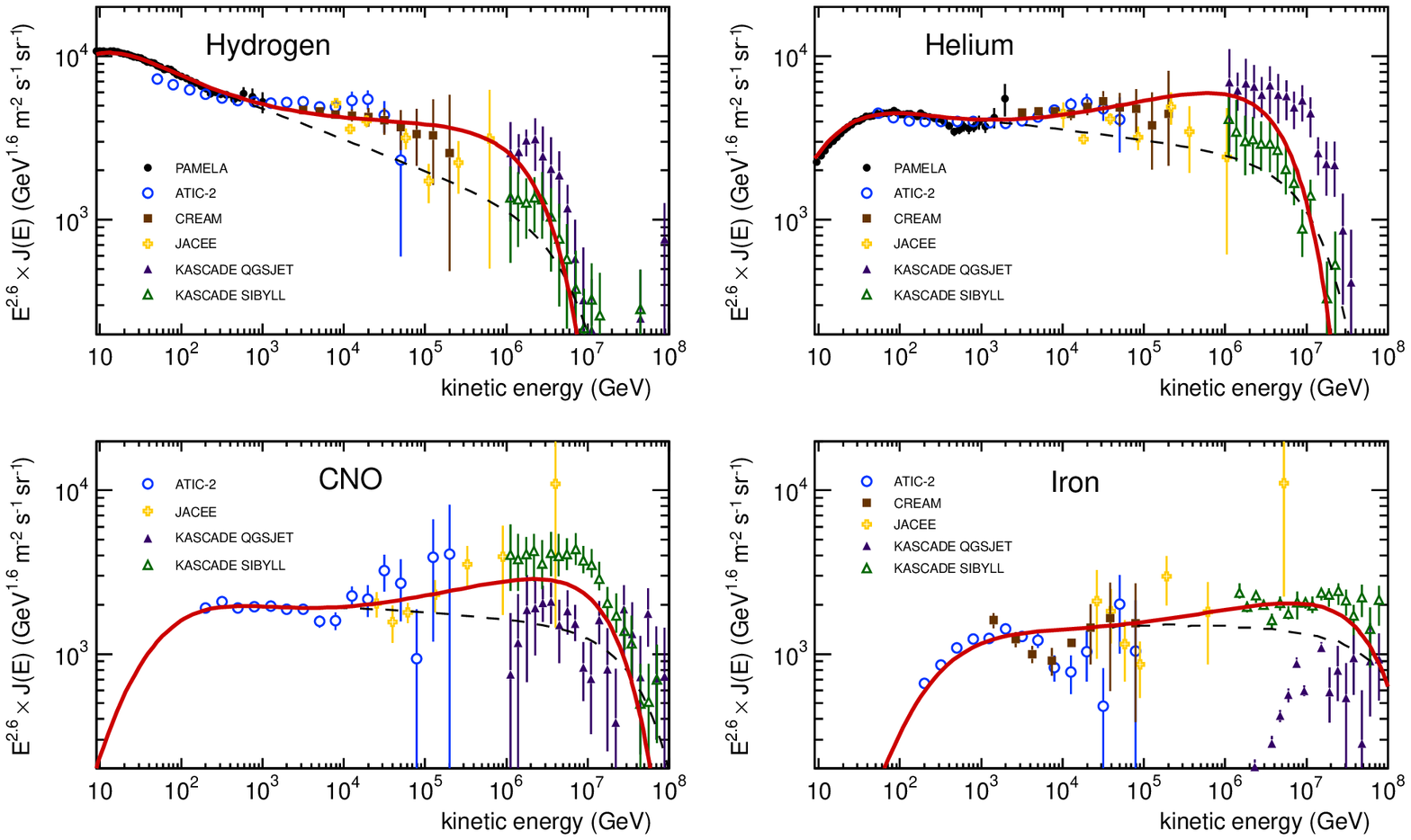}
\figcaption{ 
  CR spectra for \Hyd, \He, \C\N\Oxy{} and \Fe{} from our
  calculations and data as function of kinetic energy. The data are from 
  \PAMELA{} \citep{Adriani2011}, ATIC-2 \citep{Panov2009}, CREAM \citep{Ahn2010,Yoon2011},
  JACEE \citep{Asakimori1998}, and KASCADE \citep{Antoni2005}.
  \label{Fig::ccElementalGroups}
}
\end{center}
\end{figure*}

We now put into practice our hypothesis on the latitudinal variations of the CR diffusion properties,
which implies that $K(z,E)$ is not separable as $f(z)K_{0}(E)$ everywhere. 
We follow the arguments given in \citet{Erlykin2002}, but our 
aim is not to inspect the astrophysical plausibility of their suggestions. 
Rather, we consider a phenomenological scenario 
in order to illustrate the effect and its consequences for the main CR observables. 
We adopt a simple \textit{two-halo model} (\THM) consisting in two diffusive zones.
The \textit{inner halo},
representing the region influenced by SNRs, is taken to surround the
disk for a typical size $\xi L$ of a few hundred pc ($\xi\sim$\,0.1).
The \textit{outer halo}, representing a wider region where the turbulence is driven by CRs,
is defined by $\xi L <|z|<L$.
The diffusion coefficient is taken of the type
\begin{equation} \label{Eq::TwoHaloDiffCoeff}
K(z,\rho) = 
\begin{cases}
 k_{0}\beta\rho^{\delta} & \,{\rm\it for\,} |z|< \xi L \,\,({\rm\it inner\, halo}) \\
 k_{0}\beta\rho^{\delta+\Delta} & \,{\rm\it for\,} |z|>\xi L \,\,({\rm\it outer\, halo}) \,,
\end{cases}
\end{equation}
where $\rho=R/R_{0}$ and $k_{0}$ specifies its normalization at the reference rigidity $R_{0}$.  
We use the same $k_{0}$ values on both halos, as their relative normalization can be controlled by $R_{0}$.  
For $R>R_{0}$, the diffusion coefficient of Eq.\,\ref{Eq::TwoHaloDiffCoeff} 
produces a higher CR confinement in the inner halo (with
$\delta\sim$\,1/3), whereas the outer halo 
(with $\Delta\sim$\,0.5 -- 1) 
represents a reservoir from which CRs leak out rapidly and can re-enter the inner halo.
From Eq.\,\ref{Eq::Identity}, one can compute $\lambda=\lambda(z,\rho)$ and
$\Lambda=\Lambda(\rho)$. The latter reads
\begin{equation}\label{Eq::LvsRig}
\Lambda(\rho) = L\left[ \xi + (1-\xi)\rho^{-\Delta} \right] \,.
\end{equation}
That is, the effective halo height is a rigidity-dependent quantity
that affects the model predictions at $z=0$.
This effect can be better understood if one neglects the term $\Gamma^{\rm inel}$ and
takes a source term $Q_{\rm pri} \sim \rho^{-\nu}$. From
Eq.\,\ref{Eq::SolutionVSz} and Eq.\,\ref{Eq::LvsRig} one finds
\begin{equation}\label{Eq::SpectrumTwoComponents}
N_{0} \equiv N(z=0) \sim \frac{L }{k_{0}} \left\{ \xi \rho^{-\nu -\delta} +
  (1-\xi)\rho^{-\nu -\delta -\Delta} \right\} \,,
\end{equation}
which describes the CR spectrum as a result of two components. 
Its differential log-slope as a function of rigidity reads
\begin{equation}\label{Eq::LogSlope}
\gamma(\rho) = - \frac{d\log N_{0}}{d\log \rho} \approx \nu + \delta +
\frac{\Delta}{1 + \frac{\xi}{1-\xi}\rho^{\Delta}}  \,,
\end{equation}
which indicates a clear transition between two regimes.
In practice, the low-energy regime ($\gamma\approx\nu+\delta+\Delta$) is never reached
due to spallation (neglected in the above equations)
that becomes relevant and even dominant over escape 
($\Gamma^{\rm inel} \gtrsim \frac{K_{0}}{h\Lambda}$). In this case the log-slope  
is better approximated by $\gamma\sim \nu + \frac{1}{2}(\delta +\Delta)$ \citep{Blasi2012a}.
The hard high-energy regime ($\gamma \approx \nu+\delta$) 
is determined by the diffusion properties of the inner halo only, 
because the outer halo is characterized by a much fast
particle leakage. In this limit one has $\Lambda \approx \xi L$.
The effect vanishes at all rigidities when passing to the \HM{} limit of $\xi\rightarrow 1$ (one-halo) 
or $\Delta\rightarrow 0$ (identical halos), where one recovers the usual relation $\gamma = \nu + \delta$.
Furthermore, from Eq.\,\ref{Eq::SolutionVSz}, it can be seen that the intensity of the harder component 
diminishes gradually with increasing $|z|$, \ie, the CR spectra at
high energies are steeper in the outer halo.

\section{Results and Discussions}  
\label{Sec::Results}               

We compute the \THM{} spectra at Earth by $J(E) =\frac{\beta c}{4\pi}N_{0}(E)$, from Eq.\,\ref{Eq::SolutionVSz},
at kinetic energies above 10\,GeV.
The SNR energy spectra are taken as $Q_{\rm pri} = Y\beta^{-1}\left({R/R_{0}}\right)^{-\nu}e^{-R/R_{\rm max}}$, 
where $R_{\rm max}$ represents the maximum acceleration rigidity attainable by SNRs. 
The constants $Y$ are determined from the data at $\sim$\,100\,GeV/nucleon. 
The indices $\nu$ are taken as $Z$-dependent to account for the observed discrepancies among elements.
\citet{Malkov2012} and \citet{Ohira2011} gave strong arguments for ascribing such discrepancies to SNRs. 
The ISM surface density is taken as $h\times n\cong$\,100\,pc\,$\times$\,1\,cm$^{-3}$.
The two halos are defined by $L\cong$\,5\,kpc and $\xi L\cong$\,0.5\,kpc, 
but the physical parameters that enter the model are $k_{0}/L$ and $\xi$, 
where both quantities are also degenerated with $R_{0}$. 
Concerning the diffusion parameters, we consider two case studies:
\THMa{} ($\delta\cong$\,1/3 and $\Delta\cong$\,0.55) 
which adopts a Kolmogorov-type diffusion in the inner halo,
and \THMb{} ($\delta\cong$\,1/6 and $\Delta\cong$\,0.55) to test the extreme case of a very slow diffusivity.
The main parameters are summarized in Tab.\,\ref{Tab::ModelParameters}.
The practical implementation of the model follows \citet{NTFD2012}, see also \citet{Maurin2001}.
%
\setlength{\tabcolsep}{0.072in} 
\begin{deluxetable}{lcc}
\tablecaption{
Model parameters
  \label{Tab::ModelParameters}
}
\tablehead{
  \colhead{parameters} &  \colhead{\THMa{}} &  \colhead{\THMb{}} 
}
\startdata
$\nu$\,(H;\, He;\, CNO;\, Fe) &    2.29; 2.17; 2.17; 2.20 &    2.43; 2.31; 2.31; 2.34   \\
$R_{0}$;\,\, $R_{\rm max}$\dotfill & 2\,GV;\, 2.5$\cdot$10$^{6}$\,GV  & 2\,GV;\, 2.5$\cdot$10$^{6}$\,GV  \\
$k_{0}/L$\dotfill & 0.007\,kpc/Myr  &  0.010\,kpc/Myr  \\ 
$\delta$;\,\,$\Delta$;\,\,$\xi$\dotfill & 1/3;\,  0.55;\,  0.1  &  1/6;\,  0.55;\,  0.1 
\enddata
\end{deluxetable}

The energy spectra of \Hyd, \He, \C\N\Oxy{} and \Fe{} are shown in Fig.\,\ref{Fig::ccElementalGroups} 
in comparison with the data.
Results are shown for \THMb{} only (\THMa{} and \THMb{} predictions
are indistinguishable for primary CRs).
Our calculations (solid lines) are in good agreement with the data 
within their uncertainties. 
At low energies ($\lesssim$\,10\,GeV/nucleon) the solar modulation is apparent 
and it is described using a \textit{force-field} modulation potential $\phi\cong$\,400\,MV \citep{Gleeson1968}. 
Note, however, that our model may be inadequate in this energy region due to the low-energy approximations. 
At higher energies, our model reproduces well the observed changes in slope, 
in agreement with the trends indicated by the data.
It should be noted, however, that the sharp spectral structures suggested by the \PAMELA{}
data at $\sim$\,300\,GeV cannot be recovered. 
The \THM{} predictions are also compared with \HM{} power-law extrapolations (dashed line) 
to better illustrate the differences. 
It can be seen that the spectral upturn is slightly less pronounced for 
elements with large mass $M$ such as \Fe{},
due to the competing action of the term $\Gamma^{\rm inel}$ 
below a few 100\,GeV/nucleon (note that $\Gamma^{\rm inel}\propto M^{0.3}$).
Above $\sim$\,100\,TeV/nucleon, the elemental \textit{knees} are matched well using $R_{\rm max}=$\,2.5$\cdot$10$^{6}$\,GV, 
which is somewhat lower than that from \HM-based estimates \citep{Blasi2012a}. 
For instance, the \HM{} spectra of Fig.\,\ref{Fig::ccElementalGroups} (dashed lines) employ $R_{\rm max}=$\,7$\cdot$10$^{6}$\,GV. 

\begin{figure}
\begin{center}
\epsscale{1.15}
\plotone{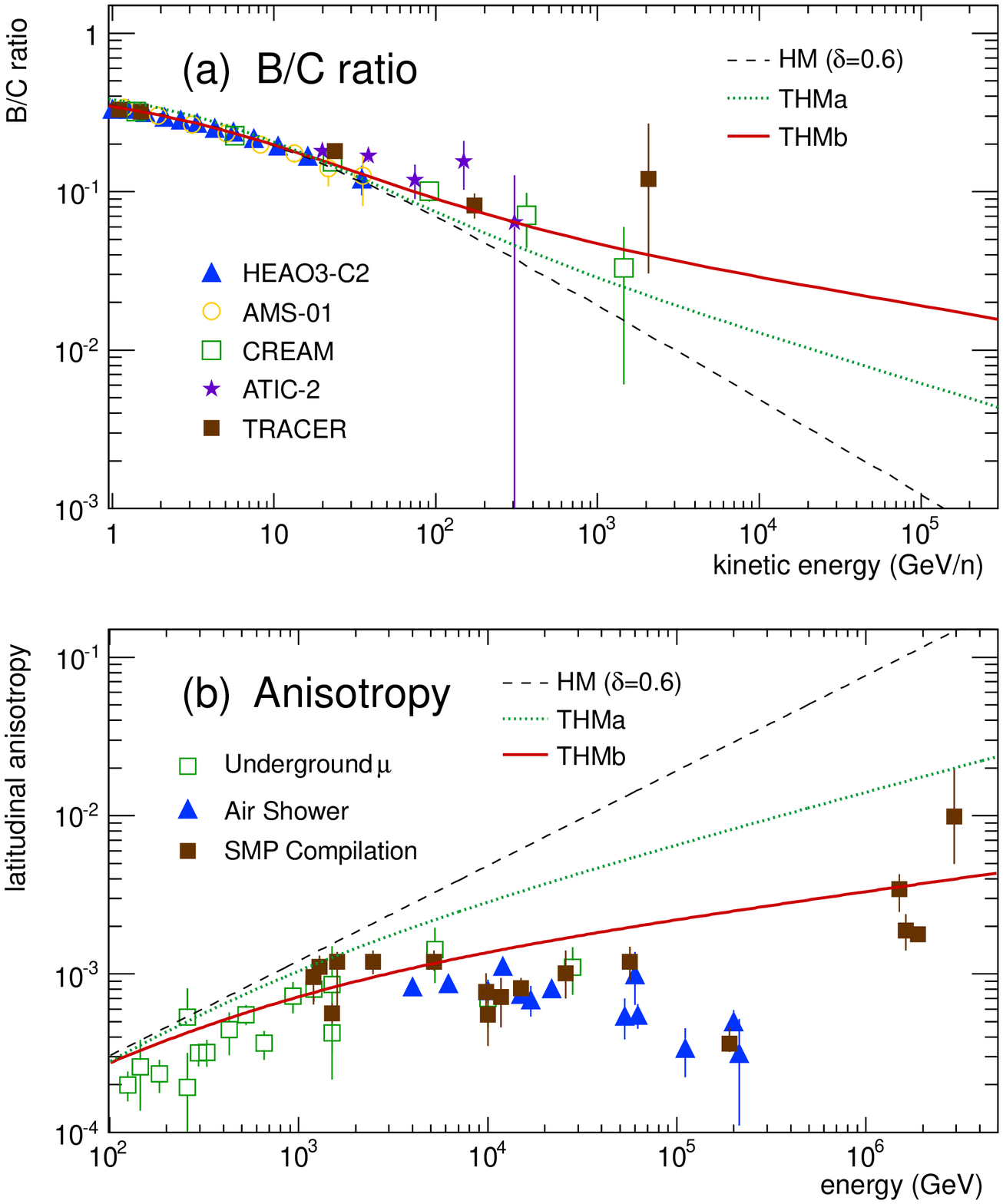}
\figcaption{ 
  \BC{} ratio (a) and latitudinal anisotropy (b) for \THMa, \THMb{} and \HM{}.
  \BC{} data are from
  HEAO3-C2 \citep{Engelmann1990}, CREAM \citep{Ahn2008}, 
  \AMS-01 \citep{AMS01Nuclei2010}, ATIC-2 \citep{Panov2007}, and TRACER \citep{Obermeier2011}.
  \textit{Air Shower} and \textit{Underground $\mu$} data are from
  \citet{Amenomori2005}; \textit{SMP} data are from \citet{Strong2007}.  
  \label{Fig::ccBCRatio}
}
\end{center}
\end{figure}

A distinctive feature of our model is provided by the secondary-to-primary ratios.
From Eq.\,\ref{Eq::SolutionVSz}, neglecting $\Gamma^{\rm inel}$, it is easy to see that, 
roughly, secondary-to-primary ratios must harden as $\propto \Lambda/K_{0}$.
In Fig.\,\ref{Fig::ccBCRatio}a we plot the \BC{} ratio. 
Remarkably, \THMb{} describes well the data down to energies of 
1\,GeV/nucleon, notwithstanding the low-energy approximations.
Its low-energy behavior is very similar to that of \HM{}, which uses $\delta=$\,0.6 in the whole halo.
At energies above $\sim$\,50\,GeV/nucleon, \THMb{} predicts a
significant flattening for the \BC{} ratio, which is also suggested
by recent data from TRACER. 
Similar conclusions can be drawn for \THMa{}.
We also expect a similar trend (but less obvious) for the \pbarp{} ratio,
which should be flatter than that estimated in \citet{Donato2011}. 
All these features can be tested by the upcoming Alpha Magnetic
Spectrometer (\AMS)\footnote{\url{http://www.ams02.org}} data on
elemental spectra (\Hyd{} to \Fe) and nuclear ratios (\BC{} or \pbarp). 
We remark that
a multi-channel study may be necessary to disentangle this 
effect from other possible processes, see, \eg, \citet{NTFD2012}. 

The hardening of the \BC{} ratio is also connected with the anisotropy.
This can be illustrated for the latitudinal component of the anisotropy,
$\eta_{z}$, which is due to the vertical outflow of CRs from the Galactic plane. It can be written as
\begin{equation}\label{Eq::Anisotropy}
  \eta_{z}(E) \approx \frac{3 K_{0}(E)}{c \Lambda(E)}\frac{z_{\odot}}{h} \,,
\end{equation}
where $z_{\odot}/h\cong$\,0.2 \citep[and references therein]{Jones2001}. 
Figure\,\ref{Fig::ccBCRatio}b shows $\eta_{z}(E)$
for the considered models compared with direct measurements of the anisotropy amplitude.
Due to limitations implicit in Eq.\,\ref{Eq::Anisotropy} as well as possible
contributions from stochasticity \citep{Blasi2012b},
the model comparison with the data is necessarily qualitative.
We focus on the comparison among the models, which is more instructive.
$\eta_{z}$ increases with energy as 
$\sim$$K_{0}/\Lambda$ (which is fixed by the
\BC{} ratio and cannot be arbitrarily changed)
so that the \HM{} yields $\eta_{z}\propto\,E^{\delta}$, whereas 
\THMa{} and \THMb{} predict a natural flattening of $\eta_{z}$. 
In particular, \THMb{} predicts a much lower anisotropy 
while keeping a good agreement with the \BC{} data at
$E\lesssim$\,100\,GeV/nucleon.
Its high-energy behavior $\eta_{z} \propto E^{1/6}$ is similar
to that of \textit{Scenario P} in \citet{Vladimirov2011}, 
but our model does not fall into any of their scenarios.
It is interesting noticing that the anisotropy may also be reduced at all energies 
if one accounts for a proper radial dependence for $K$. 
This is shown in \citet{Evoli2012}, which assumed a spatial correlation between diffusivity and SNR density.
In our \THMb{} model, the source spectra are fairly soft ($\nu\gtrsim$\,2.3) for the DSA requirements, 
however they are not exceedingly soft according to recent calculations 
where the Alfv\'en speed of the scattering centers is accounted \citep{Caprioli2012}.
On the other hand, \THMa{} gives a dependence $\eta_{z}\propto E^{1/3}$ at high energies, 
but requires $\nu\approx$\,2.2, which agrees better with the basic DSA predictions.
We recall that state-of-the-art models employ $\nu\approx$\,2.4 and $\delta\approx$\,1/3 \citep{Strong2007}
and require large amounts reacceleration to match the \BC{} data.
Besides, other \BC{}+\pbarp{} combined analyses favor
$\delta\approx$\,1/2 and smaller amounts of reacceleration \citep{DiBernardo2010}. 
Within our scenario, 
the observed steepness of the low-energy \BC{} ratio shares the same origin of
the high-energy hardening of the primary spectra. 
It can be therefore explained, under a purely diffusive picture,
why all \HM-based studies lead to systematically large values for the parameter $\delta$.

\section{Conclusions}     
\label{Sec::Conclusions}  

In this Letter, we have proposed a new interpretation for the spectral hardening observed in CRs.
As shown, it may be a consequence of a spatial change of the CR diffusion properties in the Galaxy.
From this scenario the hardening arises naturally as a local effect and vanishes gradually in the
outer halo, where the CR spectra are also predicted to be steeper.
This effect must be experienced by all CR nuclei as well as by secondary-to-primary ratios. 
Recent data (\eg, \Fermi/LAT or TRACER) seem to support this scenario,
but the predicted spectral upturn is more gradual than that suggested by \PAMELA{} data.
With dedicated analysis of the data forthcoming from \AMS{} or \Fermi/LAT,
our model can be resolutely tested and discriminated against other interpretations.

Our scenario has remarkable implications for the CR physics. 
With the \THMa{} setup, we have shown that a Kolmogorov diffusion for the inner halo ($\delta\sim$\,1/3)
is consistent with relatively hard source spectra ($\nu\sim$\,2.2).
With the \THMb{} setup, we have shown that a very slow diffusion for the inner halo
($\delta\sim$\,1/6) can reconcile the anisotropy with the \BC{} ratio.
Interestingly, it does not require prohibitively steep source spectra
as one might expect.
Both formulations are based on plain diffusion models where the diffusion coefficient is not 
separable into energy (rigidity) and space terms as usually assumed.
Their good agreement with data suggests that the CR propagation at $\lesssim$\,50\,GeV/nucleon 
might be only moderately affected by low-energy effects such as reacceleration. 
At the highest energies, we found that the elemental knees can be matched using  
$\sim$\,2.5$\cdot$10$^{6}$\,GV of maximum SNR rigidity, which is 
attainable by known acceleration mechanisms. 

Further elaborations may be performed using numerical methods, 
in order to introduce a radial dependence of $K$ or other effects
connected with the Galactic structure. Calculations of other
observables such as \pbarp, $^{10}$Be/$^{9}$Be, or  $\gamma$--rays 
may lead to a deeper understanding of the effect.
We hope that our proposal will stimulate further investigations on this subject. 
\\
I thank F. Donato for insightful suggestions and B. Bertucci for constant encouragement. 
This work is supported by the Italian Space Agency under contract ASI-INFN I/075/09/0.
\\


\end{document}